\documentclass[aps,reprint,twocolumn,preprintnumbers,amsmath,amssymb,nofootinbib,superscriptaddress,showkeys]{revtex4}
\usepackage{lipsum}
\usepackage{graphicx}
\usepackage{epsfig}
\usepackage{epstopdf}
\usepackage{hyperref}
\usepackage{amsmath}
\usepackage{amsfonts}
\usepackage{amssymb}
\usepackage{setspace}
\usepackage{amsfonts,color}
\usepackage{verbatim}

\usepackage{color}

\def\be{\begin{equation}}
\def\ee{\end{equation}}
\def\bea{\begin{eqnarray}}
\def\eea{\end{eqnarray}}

\begin{document}

\title{Open-string singlet interaction as pomeron in elastic $pp, p\bar{p}$ scattering}

\author{Piyabut Burikham}
\email{piyabut@gmail.com}

\author{Daris Samart}
\email{dsamart82@gmail.com}

\affiliation{High Energy Physics Theory Group, Department of Physics, Faculty of Science, Chulalongkorn University, Phyathai Rd., Bangkok 10330, Thailand}

\date{\today}

\begin{abstract}
We consider contributions of the open-string singlet~(OSS) interaction to the proton-proton scattering. Modelling baryon and meson as instanton and open string in the effective QCD string model, the typical pomeron is identified with a massive spin-2 closed string. According to the open-closed string duality, the massive spin-2 closed string is accompanied by the open-string singlet interaction with universal coupling.  The flat space 4-point open-string singlet amplitudes and cross sections are computed in the Regge limit.  Including both the open and closed string pomerons, we fit the differential cross-section with experimental data of the proton-proton and proton-antiproton scattering from the CDF, E-710, D0, LHC-ATLAS and LHC-TOTEM collaborations. The fitting results show the universality of the string mass scale~(or string tension) and consistency of the string model between the open and closed string parameters in describing the $pp$ and $p\bar{p}$ scattering in the Regge limit. The fitting of the string model to experimental data gives the string scale $M_{S}\simeq 1.7-1.8$ GeV and the corresponding glueball pole mass $m_{g}\simeq 1.6-1.7$ GeV. This lies within experimental values of the observed $2^{++}, I=0$ ``meson'' masses.

\end{abstract}

\maketitle
\section{Introduction}

String theory first appeared as the physical interpretation of the Veneziano amplitude describing meson scatterings~\cite{Nambu:1969se,Koba:1969rw,Susskind:1970qz}. Tower of resonance states with increasing masses continuing indefinitely explains the zoo of hadronic particles found and at the same time quantitatively accounts for their high energy scattering in the so-called Regge limit~(large center of mass energy and small transfered momentum).  The string picture of strong interaction eventually fell out of favour due to the success of the Quantum Chromodynamics~(QCD) in the description of the Deep Inelastic scattering~(DIS) between leptons and hadrons confirming existence of fractionally charged quarks.  QCD with quark model can also describe all known hadronic states and predicts more exotic ones already observed~\cite{Tanabashi:2018oca}.  Moreover, the strongest motivation of string theory as fundamental theory of quantum gravity, originated from the inevitable prediction of the massless spin-2 states that couple universally to other particles~\cite{Yoneya:1974jg,Scherk:1974ca} as a result of the open-closed string duality~\cite{Green:1987sp}, was ironically the crucial failure of the string theory as the fundamental framework of the strong interaction. Experimentally, there is no massless spin-2 hadronic states.

Despite the success of QCD in the perturbative regime, it fails to give accurate results on the diffractive scattering~(mostly small momentum transfer $t$) and the total cross section of the high-energy hadronic scattering.  Experiments found the increase of the total cross sections of the hadronic processes with respect to the scattering energy~\cite{Donnachie:1992ny}.  Regge theory provides explanation of the rise of the forward amplitude and the total cross section by proposing a trajectory of states with angular momenta $J=\alpha_{0}+\alpha' t$ with $\alpha_{0}\gtrsim 1$, the so-called pomeron trajectory.  In the crossing channel, the trajectory relates the angular momentum of the resonance states with the mass square $J=\alpha_{0}+\alpha' M^{2}$.  The pomeron trajectory contains states with vacuum quantum number but the masses are unknown. From the experimental data fittings, it is very likely that the lowest mode of pomeron is the state with spin-2~\cite{Ewerz:2016onn}.  In this Regge regime~(at high energies and small $t$, notably the perturbative QCD on the other hand, is applicable for the high-energy scattering with large momentum transfer, i.e. large $t$), the string description as a flat space realization of the S-matrix formalism of the Regge theory is a better framework to address the scattering processes.

Lattice QCD actually predicts existence of massive spin 0, 1 and 2 colour-singlet glueball states~\cite{Ochs:2013gi}. The conformal symmetry is dynamically broken at low energies by the quantum fluctuations and mass gap is generated.  Naturally, the spin-2 glueball state is the best candidate of the pomeron with other spin-0,1 states responsible for the daughter trajectories.  Revival of the string description of the strong interaction came with the proposal of the AdS/CFT correspondence~\cite{Maldacena:1997re} where it was argued that strongly interacting gauge theory living on the AdS boundary is dual to the weakly coupled gravity in the AdS bulk.  Notably, the gauge theory on the boundary does not have gravitons but the correlation $<T_{\mu\nu}T^{\mu\nu}>$ can be generated via the exchanges of (massive) gravitons in the bulk.  In this way, the problem of existence of massless spin-2 particle in the boundary theory is resolved.  Meson as the quark-antiquark bound state can be represented by an open string hanging in the bulk whose ends are living on the boundary.  Baryon can be modelled as the open-string instanton with three open strings joining at a baryon vertex with another end of each string living on the boundary~\cite{Witten:1998xy,Gross:1998gk}. Generalized multiquark states can naturally be described as systems of connecting open strings~\cite{Burikham:2008cr}. Scattering of mesons and baryons thus can be thought of as scattering of open strings living close to the AdS boundary.  For high energy scattering, each quark effectively appears as open string hanging all the way down into the bulk with only one end living on the boundary and the scattering can be described by the perturbative QCD.

In this unifying framework of ``holographic QCD''~(or ``AdS/QCD''), apart from the missing details of mass-gap generating mechanism, all of the strong interaction processes could be addressed at least in principle~(or if ones do not want to take perturbative QCD as a limit of the holographic gauge theory in the string setup, ones can think of the two theories as complementary to one another in describing the strong interaction).  The glueball masses or mass gap can be introduced by inserting a cutoff in the AdS coordinate, breaking conformal symmetry of the gauge theory on the boundary.  The exchange of glueball as pomeron on the boundary is dual to the exchange of massive graviton~(closed string) in the bulk.  In this way, glueball is a closed string pomeron.  On the other hand, the open-closed string duality implies existence of open-string singlet interaction~\cite{Burikham:2006an} between gauge singlet states.  The graviton exchange diagram of the $2\to 2$ process is the nonplanar annulus worldsheet with 2 vertex operators inserted at each boundary~\cite{Green:1987sp}. This diagram is dual to the sum of one-loop open-string diagrams containing two twists.  Since even the singlet states couple to closed-string graviton via the energy-momentum tensor, they must also inevitably couple through open-string singlet.  Interestingly, the 4-point amplitude for the open-string singlets are non-vanishing even though the 3-point is~\cite{Burikham:2006an,Cullen:2000ef,pbthesis,Burikham:2003ha,Burikham:2004su,Burikham:2004uu,Burikham:2006hi}.  Moreover, the duality of the open-string singlet amplitude naturally eradicates the zeroth mode~(as well as the even modes) of the string resonances and its Regge limit behaviour fits excellently with the pomeron as we will see in subsequent sections.

A study of the $pp, p\bar{p}$ scattering in the Regge limit has been investigated in Ref. \cite{Domokos:2009hm} by using the AdS/QCD framework and its consequences for other scattering processes~\cite{Domokos:2010ma,Avsar:2009hc,Anderson:2014jia,Iatrakis:2016rvj,Anderson:2016zon,Hu:2017iix,Xie:2019soz}. Authors in Ref. \cite{Domokos:2009hm} have used massive graviton~(closed-string) as exchange particle of the scattering at the tree-level and identified the massive graviton as tensor pomeron or glueball. Although results of the model parameters from AdS/QCD calculations in Ref.~\cite{Domokos:2009hm,Domokos:2010ma} are in good agreement with the fitting results from the experimental data. However, one must also take into account the effect of open-string singlet interaction in the scattering instead of using only the closed-string exchange. In this work, we investigate the inclusion of open-string pomeron contribution in the $pp, p\bar{p}$ elastic scattering in the Regge regime.  In Section~\ref{Form}, we calculate the scattering amplitude of the OSS interaction for proton-proton scattering in the Regge limit. Fitting results of the model parameters will be determined in Section~\ref{Fitt}. Section~\ref{Disc} discusses the results and concludes the work.

\section{Formalism}  \label{Form}
\subsection{Scattering amplitude from the open-string singlet interaction }

We start with the OSS amplitude for elastic proton-proton scattering. In holographic QCD, proton can be described by three open strings joining at a baryon vertex and hanging close to the boundary of an asymptotically AdS space. We can imagine when two protons collide, a string from each baryon vertex will scatter off one another.  Instead of considering detailed scattering dynamics of the strings and smeared baryon vertices in the curved AdS space and also motivated by success of the dual resonance model in the Regge regime, we will assume the scattering can be described by simple flat-space open-string amplitude. The general form of the four-fermion interaction of the open string can be written as~\cite{Burikham:2003ha}
\begin{eqnarray}
\mathcal{A}_{\rm string} &=& - g_S\left( t\,\bar\psi_2\gamma_\mu\psi_1\,\bar\psi_4\gamma^\mu\psi_3 - s\,\bar\psi_4\gamma_\mu\psi_1\,\bar\psi_2\gamma^\mu\psi_3 \right)
\nonumber\\
&&\times\Bigg( \frac{S(s,\,t)}{s\,t}\,T(1234)
\nonumber\\
&&\qquad +\, (1\leftrightarrow 4,\,s\leftrightarrow u) + (1\leftrightarrow 2,\,t\leftrightarrow u)\Bigg),
\label{4-fermion-open-string}
\end{eqnarray}
where $g_S$ is the dimensionless string coupling and the particles (fermions, $\psi_i$) in the two-body scattering are labeled by $12\,\to\,34$\,. The function $S(x,\,y)$ is the Veneziano-like amplitude defined by \cite{Veneziano:1968yb}
\begin{eqnarray}
S(x,y) = \frac{\Gamma\left(1-\alpha'x\right)\,\Gamma\left(1-\alpha'y\right)}{\Gamma\left(1-\alpha'x -\alpha'y \right)}\,,
\end{eqnarray}
where $\alpha' \equiv M_S^{-2}$ and $M_S$ is the string mass scale. The $T(ijkl)$ are the Chan-Paton factors of the matrices $\lambda^{a}\in U(n)$ and they are defined by
\begin{eqnarray}
T(1234) = {\rm tr}(\lambda^1\lambda^2\lambda^3\lambda^4) + {\rm tr}(\lambda^4\lambda^3\lambda^2\lambda^1)\,,
\end{eqnarray}
with the normalization ${\rm tr}(\lambda^a\,\lambda^b) = \delta^{ab}$. The Mandelstam variables are defined as
\begin{eqnarray}
s = -(p_1 + p_2)^2,\; t = -(p_1 + p_4)^2, \; u = -(p_1 + p_3)^2,
\end{eqnarray}
where all momenta are directed inward. As shown in Ref. \cite{Cullen:2000ef,Burikham:2006an}, the stringy gauge singlet or the OSS interaction occurs between all gauge singlet particles in the $U(n)$ group that give $T(1234)=T(1324)=T(1243)\equiv T$\,. One then can rewrite the four-fermion amplitude for the OSS as
\begin{eqnarray}
\mathcal{A}_{\rm OSS} &=& -\,g_S\,T\left( t\,\bar\psi_2\gamma_\mu\psi_1\,\bar\psi_4\gamma^\mu\psi_3 - s\,\bar\psi_4\gamma_\mu\psi_1\,\bar\psi_2\gamma^\mu\psi_3 \right)
\nonumber\\
&&\times\,\frac{1}{s\,t\,u}\,f(s,t,u)\,,
\end{eqnarray}
where the $f(s,\,t,\,u)$ function is
\begin{eqnarray}
f(s,t,u) = u\,S(s,t) + t\,S(s,u) + s\,S(u,t)\,.
\end{eqnarray}
Applying the Fierz transformation~\cite{Itzykson:1980rh}, we obtain
\begin{eqnarray}
4\bar\psi_i\gamma_\mu\psi_j\,\bar\psi_k\gamma^\mu\psi_l &=& 4\,\bar\psi_k\psi_j\,\bar\psi_i\psi_l
- 2\,\bar\psi_k\gamma_\mu\,\gamma_5\psi_j\,\bar\psi_i\gamma^\mu\,\gamma_5\psi_l
\nonumber\\
&-& 2\,\bar\psi_k\gamma_\mu\psi_j\,\bar\psi_i\gamma^\mu\psi_l
- 4\,\bar\psi_k \gamma_5\psi_j\,\bar\psi_i\gamma_5\psi_l .
\nonumber\\
\end{eqnarray}
Then, the $\mathcal{A}_{\rm OSS}$ amplitude reads,
\allowdisplaybreaks
\begin{widetext}
\begin{eqnarray}
\mathcal{A}_{\rm OSS} &=& -\frac{1}{4}\,g_S\,T\Big[\,  t\left( 4\,\bar\psi_4\psi_1\,\bar\psi_3\psi_2
- 2\,\bar\psi_4\gamma_\mu\,\gamma_5\psi_1\,\bar\psi_3\gamma^\mu\,\gamma_5\psi_2 - 2\,\bar\psi_4\gamma_\mu\psi_1\,\bar\psi_3\gamma^\mu\psi_2
- 4\,\bar\psi_4 \gamma_5\psi_1\,\bar\psi_3\gamma_5\psi_2 \right)
\\
&& \qquad\quad -\, u\left( 4\,\bar\psi_3\psi_1\,\bar\psi_4\psi_2
- 2\,\bar\psi_3\gamma_\mu\,\gamma_5\psi_1\,\bar\psi_4\gamma^\mu\,\gamma_5\psi_2 - 2\,\bar\psi_3\gamma_\mu\psi_1\,\bar\psi_4\gamma^\mu\psi_2
- 4\,\bar\psi_3 \gamma_5\psi_1\,\bar\psi_4\gamma_5\psi_2  \right) \Big] \frac{1}{s\,t\,u}\,f(s,t,u)\,.
\nonumber
\end{eqnarray}
\end{widetext}
Note that the amplitude is naturally Reggeized by construction. To see the Regge behaviour in the limit $s$ large and fixed small $t$, we recall some relevant properties of the gamma function,
\begin{eqnarray}
\Gamma(1-x)  &=& \frac{\pi}{\Gamma(x)\,\sin(\pi\,x)}\,,\label{gamma1}
\\
\frac{\Gamma(x + a)}{\Gamma(x + b)}&\stackrel{{\rm large}~x}{\simeq}& x^{a-b}\left( 1 + \mathcal{O}\left(\frac{1}{x} \right)\right)
\label{gamma2} .
\end{eqnarray}
In case of $|s|\to \infty$ along the complex plane on the real axis of $s$, we will encounter an infinite number of poles in the $S(s,t)$ function. To avoid singularities, we consider the imaginary part of the $s$ variable by adding $i\epsilon$ to skip the poles on the real axis, this gives
\begin{eqnarray}
S(s,t) &\stackrel{{\rm large}~s}{\simeq}&
\Gamma(1-\alpha' t)\,e^{-i\pi\alpha't}\,(\alpha' s)^{\alpha' t}\,.
\end{eqnarray}
By using the same trick for $S(t, u)$, we obtain
\begin{eqnarray}
S(t, u) &\stackrel{{\rm large}~s}{\simeq}&
\Gamma(1 -\alpha' t)\,(\alpha' s)^{\alpha' t}\,.
\end{eqnarray}
Lastly, the $S(s,u)$ function in the Regge limit reads,
\allowdisplaybreaks
\begin{eqnarray}
S(s,u) &\stackrel{{\rm large}~s}{\simeq}& 0\,.
\end{eqnarray}
This term rapidly vanishes for $s\to \infty$ and there is no pole in the $t$ channel. We refer detailed derivation of the Veneziano amplitude in the Regge limit in \cite{Frampton:1986wv,Donnachie:2002en,Wong:1995jf}.
The amplitude function $f(s,t,u)$ in the Regge limit is then
\begin{eqnarray}
f(s,t,u) &=& u\,S(s,t) + t\,S(s,u) + s\,S(u,t)
\label{fstu-regge}
\\
&\stackrel{{\rm large}~s}{\simeq}& \frac{1}{\alpha'}\,\Gamma(1-\alpha' t)\left(1- e^{-i\pi\alpha't}\right)(\alpha' s)^{\alpha' t + 1} \,,
\nonumber
\end{eqnarray}
where $u \simeq - s$ has been used in the Regge limit.  The stringy $s,t,u$ symmetric function in the amplitude becomes
\begin{eqnarray}
\frac{f(s,t,u)}{s\,t\,u} &\to& -\frac{\alpha'}{t}\,\Gamma(1-\alpha' t)\left(1- e^{-i\pi\alpha't}\right)(\alpha' s)^{\alpha' t - 1}\,.
\nonumber\\
\end{eqnarray}
Finally, the OSS amplitude of proton-proton scattering, $\mathcal{A}_{\rm OSS}$ in the Regge limit is given by
\begin{eqnarray}
\mathcal{A}_{\rm OSS} &=& \frac{1}{4}\,g_S\,T\,\frac{\alpha'}{t}\,\Gamma(1-\alpha' t)\left(1- e^{-i\pi\alpha't}\right)(\alpha' s)^{\alpha' t - 1}
\nonumber\\
&\times&\Big[ t\big( 4\,\bar\psi_4\psi_1\,\bar\psi_3\psi_2
- 2\,\bar\psi_4\gamma_\mu\,\gamma_5\psi_1\,\bar\psi_3\gamma^\mu\,\gamma_5\psi_2
\nonumber\\
&& \quad\,-\,2\,\bar\psi_4\gamma_\mu\psi_1\,\bar\psi_3\gamma^\mu\psi_2
- 4\,\bar\psi_4 \gamma_5\psi_1\,\bar\psi_3 \gamma_5\psi_2  \big)
\nonumber\\
&& \,-\, u\left( 3\leftrightarrow4 \right) \Big]\,.
\end{eqnarray}
In $pp\to pp$ elastic process at high energies, the $t$-channel reggeon-pomeron exchange is dominant. To take into account effects of the internal structure of the proton, we introduce the dipole form-factors, $A(t)$~\cite{Pagels:1966zza, Domokos:2009hm} phenomenologically to the vector-vector amplitude and the effective proton-proton-Reggeon vertex in the following form
\begin{eqnarray}
A(t) =\left(\frac{1}{1 - t/\Lambda^2}\right)^2\,.
\end{eqnarray}
The $\Lambda$ parameter is the IR cutoff for the amplitude and we will treat it as a free parameter in this work. The final form of OSS amplitude in the Regge limit reads,
\begin{eqnarray}
\mathcal{A}_{\rm OSS} &=& \frac{1}{4}\,g_S\,T\,\frac{\alpha'}{t}\,\Gamma(1-\alpha' t)\left(1- e^{-i\pi\alpha't}\right)(\alpha' s)^{\alpha' t - 1}
\nonumber\\
&\times&\Big[ t\big( 4\,\bar\psi_4\psi_1\,\bar\psi_3\psi_2
- 2\,\bar\psi_4\gamma_\mu\,\gamma_5\psi_1\,\bar\psi_3\gamma^\mu\,\gamma_5\psi_2
\nonumber\\
&& \quad\,-\,2\,\bar\psi_4\gamma_\mu\psi_1\,\bar\psi_3\gamma^\mu\psi_2
- 4\,\bar\psi_4 \gamma_5\psi_1\,\bar\psi_3 \gamma_5\psi_2  \big)
\nonumber\\
&& \,-\, u\left( 3\leftrightarrow4 \right) \Big]\left(\frac{1}{1 - t/\Lambda^2}\right)^4.
\label{SGS-amp}
\end{eqnarray}
There are three free parameters in the amplitude of the OSS interaction for proton-proton scattering; the open-string coupling~($g_S$), the string mass scale~($M_S = 1/\sqrt{\alpha'}$) and the dipole IR-cutoff~($\Lambda$). As shown in Appendix \ref{app}, the OSS cross sections of the $pp$ and $p\bar{p}$ collisions are identical. Therefore we will evaluate these parameters by fitting with the $pp$ and $p\bar{p}$ scattering data from several collaborations in subsequent sections. To prepare observable for determining the parameters of the model from the experimental data, we will calculate the differential cross-section as function of $t$ in the next section.

\subsection{Differential cross-section}

In this section, we calculate the absolute square of the amplitude $\big| \mathcal{A}_{\rm OSS} \big|^2$ analytically by using Feyncalc \cite{Shtabovenko:2016sxi,Mertig:1990an} package of MATHEMATICA. The unpolarized spin sum of the amplitude, $\big| \mathcal{A}_{\rm OSS} \big|^2$ is given by,
\begin{eqnarray}
\sum_{\rm spin} \big| \mathcal{A}_{\rm OSS} \big|^2 %&=& \fixme{{\rm Tr}}\,\big| \mathcal{A}_{\rm SGS} \big|^2
%\nonumber\\
&=& 8\, g_S^2\, T^2\,\frac{\alpha'^2}{t^2}\left|1- e^{-i\pi\alpha't}\right|^2
\nonumber\\
&\times&\Gamma^2(1-\alpha't)\,(\alpha's)^{2\alpha't - 2}
\,A^4(t)
\nonumber\\
&\times&\Big(8 m^4 (7 s^2 + 7 s t + 5 t^2) - 8 m^2 s^3
\nonumber\\
&\;& \quad +\, 2 s^4 + 4 s^3 t + 2 s^2 t^2 + t^4 \Big)\,,
\end{eqnarray}
where $m$ is proton mass. In the $s\gg t,\,m$ limit, the $s^4$ term dominates in the last bracket and we have
\begin{eqnarray}
\frac{1}{4} \sum_{\rm spin} \big| \mathcal{A}_{\rm OSS} \big|^2
&\stackrel{s\gg t}{\simeq}& 8\, g_S^2\, T^2 \,(1-\cos(\pi\alpha't))
\label{absamp}\\
&\times& \Bigg( \frac{\Gamma(1-\alpha't)}{\alpha't}\Bigg)^2(\alpha's)^{2\alpha't + 2}\,A^4(t).   \nonumber
\end{eqnarray}
The differential cross-section $d\sigma/dt$ is given by
\begin{eqnarray}
\frac{d\sigma}{dt} &=& \frac{1}{16{{\pi}}\,s^2}\left(\frac14 \sum_{\rm spin} \left|\mathcal{A}_{\rm OSS}\right|^2 \right)
\nonumber\\
&=& \frac{1}{2\pi}\,g_S^2\,T^2\,\alpha'^2 \,(1-\cos(\pi\alpha't))
\nonumber\\
&&\times \Bigg( \frac{\Gamma(1-\alpha't)}{\alpha't}\Bigg)^2(\alpha's)^{2\alpha't }\,A^4(t)\,.
\label{fit-formular}
\end{eqnarray}
Finally, one obtains the differential cross-section formula of the OSS model for $pp, p\bar{p}$ scattering and we will use the formula in Eq. (\ref{fit-formular}) to fit with the experimental data.

\section{Fitting model parameters with experimental data}  \label{Fitt}
\subsection{Fitting the parameters}

We will use the experimental data from the Measurement of Elastic and Total Cross Sections in $pp$ and $p\bar p$ interactions from The Durham Hep database \cite{durham}. The data for the first fitting is from the E-710 collaboration at $\sqrt{s}= 1800$ GeV~\cite{Amos:1990fw}. The fitting to the data at other scattering energies will be presented in subsequent section.

To convert to correct units, we multiply $(0.3894)^2$ to the right-hand side of Eq. (\ref{fit-formular}) since $1$ GeV$^{-2}$ = $0.3894$ mbarn and all of phase convention from \cite{Tanabashi:2018oca} is implied. We use input values of the other parameters in Eq. (\ref{fit-formular}) as
\begin{eqnarray}
m = 0.938\;{\rm GeV}\,,\qquad \sqrt{s} = 1800\;{\rm GeV}\,,\qquad T=1\,.
\label{input-parameters}
\end{eqnarray}

\begin{figure}[t]
\centering
\includegraphics[width=.5\textwidth]{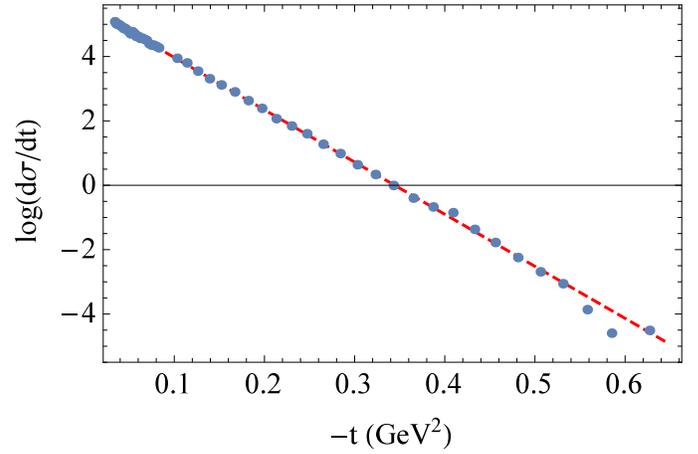}
\caption{A log-linear plot of the differential cross section at $\sqrt{s}$ = 1800 GeV from the E-710 collaboration, as a function of $-t$ (in GeV$^2$). Dotted line is the experimental data. Red line is Eq. (\ref{fit-formular}) with $g_S = 88.316$\,, $M_S = 1/\sqrt{\alpha'} = 1.351$ GeV\,, $\Lambda = 6.570$ GeV and input parameters in Eq. (\ref{input-parameters}). }
\label{plot1}
\end{figure}
Fitting the data with (\ref{fit-formular}) gives:
\begin{eqnarray}\label{fit1}
g_S = 88.316\,,\quad M_S = 1.351\;{\rm GeV},\quad \Lambda = 6.570\;{\rm GeV}.
\end{eqnarray}
In addition, if we perform the fitting by dropping the form factor i.e. $A(t)=1$. The fitting results are:
\begin{eqnarray}
g_S = 87.400,\qquad\qquad M_S = 1.344\;{\rm GeV}.
\end{eqnarray}
The plot of the best-fit parameters (\ref{fit1}) is shown in Fig. 1. We found that the fitting plots with and without the form-factor are almost identical. We can examine this situation by considering the logarithm of the differential cross section in Eq. (\ref{fit-formular}).  Considering the series expansion of Eq. (\ref{fit-formular}) and keeping $t$ at leading order, one gets
\begin{eqnarray}
\log\left(\frac{d\sigma}{dt}\right)
&=& \log\Bigg(\frac{1}{2\pi}\,g_S^2\,T^2\,\alpha'^2 \,(1-\cos(\pi\alpha't))
\nonumber\\
&&\times \Bigg( \frac{\Gamma(1-\alpha't)}{\alpha't}\Bigg)^2(\alpha's)^{2\alpha't }\,A^4(t)\Bigg)
\nonumber\\
&\approx& m_1\,t + C_1\,,
\end{eqnarray}
where
\begin{eqnarray}
m_1 &=& 2\alpha'\log(\alpha's) + 2\gamma_E\alpha' + \frac{8}{\Lambda^2} \,,
\label{slope-open}\\
C_1 &=& \log\left(\frac{\pi}{4}\,g_S^2\,T^2 \alpha'^{2}\right),
\label{intercept-open}
\end{eqnarray}
and $\gamma_E \simeq 0.5772$ is the Euler's constant. Here $m_1$ and $C_1$ are the slope and vertical intercept of the log-linear plot respectively. In the Regge limit i.e. large $s$, the slope $m_1$ can be approximated for large $\Lambda \gg 1$ as
\begin{eqnarray}
m_1 &\approx& 2\alpha'\log(\alpha's),
\label{slope-open-regge}
\end{eqnarray}
mostly insensitive to $\Lambda$. However, for small $\Lambda\simeq 1$ GeV, the slope is affected significantly from the term $8/\Lambda^{2}$ as will be shown in the next section.

\subsection{Universality of the string mass scale in the OSS interaction}

In the last section, we have performed the fitting of elastic proton-proton scattering using OSS interaction in the Regge limit. Next, we would like to see if the string mass scale $M_{S}$ is universal for other sets of the experimental data at high energies. We will use data from the LHC-ATLAS \cite{Aad:2014dca}, D0 \cite{Abazov:2012qb} and CDF \cite{Abe:1993xx} collaborations at $\sqrt{s} =$ 7000, 1960, 1800 and 546 GeV, respectively.

\begin{figure*}[t]
\centering
	\includegraphics[width=1.0\textwidth]{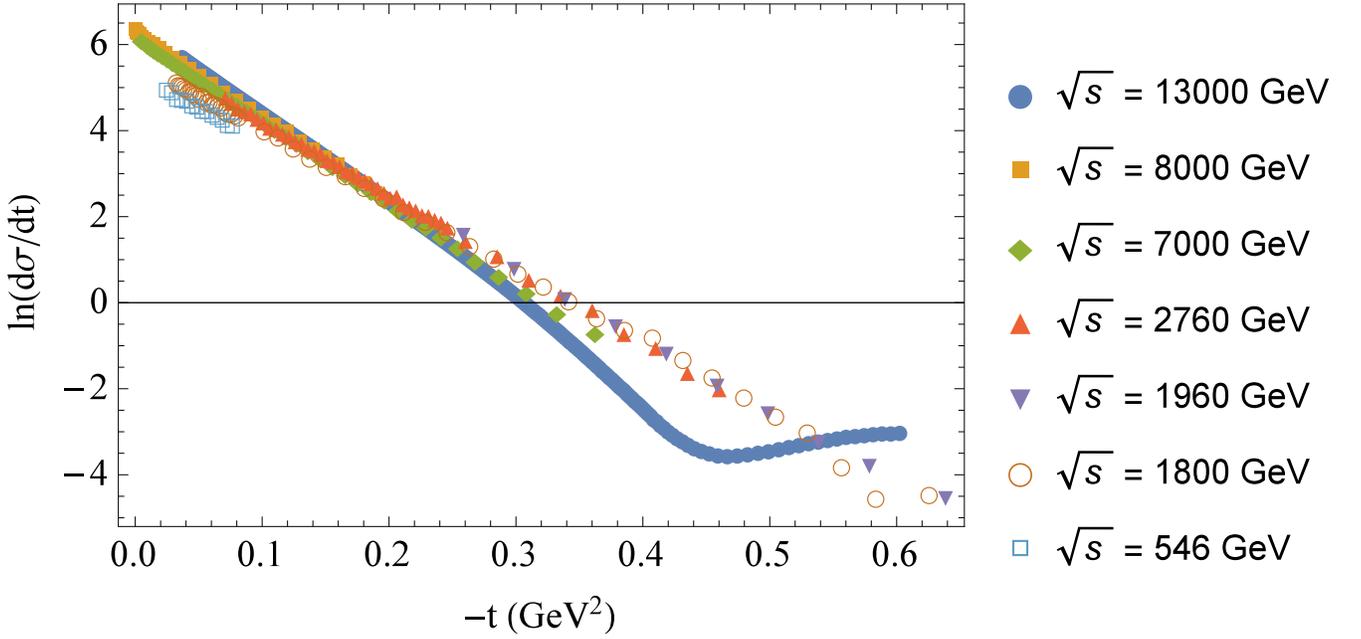}
\caption{A log-linear plot of the differential cross section as a function of $-t$ (in GeV$^2$) at $\sqrt{s}$ = 13000, 8000, 7000, 2760, 1960, 1800 and 546 GeV from the LHC-TOTEM \cite{Antchev:2018rec}, LHC-ATLAS \cite{Aad:2014dca}, D0 \cite{Abazov:2012qb}, E-710 \cite{Amos:1990fw} and CDF \cite{Abe:1993xx} collaborations. }
	\label{all-data}
\end{figure*}

In the following, the input parameters are fixed to
\begin{eqnarray}
m = 0.938\;{\rm GeV}\,,\quad \Lambda = 1.17\;{\rm GeV}\,,\quad T=1\,,
\end{eqnarray}
and consider scattering at  $\sqrt{s} =$ 7000, 1960, 1800 and 546 GeV. The value $\Lambda = 1.17$ GeV is estimated by using the Skyrme model~\cite{Cebulla:2007ei} and it is a good approximation up to $|t|< 0.8$ GeV$^2$~\cite{Domokos:2009hm}. More importantly, the data will be fit in the log-linear regime i.e., only up to $-t \simeq$ 0.6 GeV$^2$ for $\sqrt{s}$ = 1960 GeV as shown in Fig.~\ref{all-data}.

The fitting is performed in two cases; with the dipole form-factor $A(t) = 1/(1-t/\Lambda^2)^2$ and without form-factor i.e., $A(t) = 1$.

With the dipole form-factor, the fitting results are given by
\begin{eqnarray}
g_S &=& 156.887\,,\quad M_S = 1.576\;{\rm GeV}\,,  \quad \sqrt{s} = 7000 \;{\rm GeV}\,,
\nonumber\\
g_S &=& 138.728\,,\quad M_S = 1.576\;{\rm GeV}\,, \quad \sqrt{s} = 1960 \;{\rm GeV}\,,
\nonumber\\
g_S &=& 125.535\,,\quad M_S = 1.585\;{\rm GeV}\,,  \quad \sqrt{s} = 1800 \;{\rm GeV}\,,
\nonumber\\
g_S &=& 100.867\,,\quad M_S = 1.564\;{\rm GeV}\,,  \quad \sqrt{s} = 546 \;{\rm GeV}\,. \nonumber \\
\end{eqnarray}

Without form-factor, the fitting results are given by
\begin{eqnarray}
g_S &=& 114.452\,,\quad M_S = 1.355\;{\rm GeV}\,,  \quad \sqrt{s} = 7000 \;{\rm GeV}\,,
\nonumber\\
g_S &=& 90.441\,,\quad M_S = 1.357\;{\rm GeV}\,,  \quad \sqrt{s} = 1960 \;{\rm GeV}\,,
\nonumber\\
g_S &=& 87.400\,,\quad M_S = 1.344\;{\rm GeV}\,,  \quad \sqrt{s} = 1800 \;{\rm GeV}\,,
\nonumber\\
g_S &=& 66.574\,,\quad M_S = 1.272\;{\rm GeV}\,,  \quad \sqrt{s} = 546 \;{\rm GeV}\,. \nonumber \\
\end{eqnarray}
According to the results in both cases, the values of the string mass, $M_S$ are almost identical within each case. In contrast to the previous section where both cases give roughly the same fitting values of $g_{S}$ and $M_{S}$, the fitting values of $M_S$ for both cases are different by roughly 15\% with $\Lambda = 1.17$ GeV. When $\Lambda \lesssim 1$, the term $8/\Lambda^2$ in Eq. (\ref{slope-open}) becomes larger and contributes more to the slope of the differential cross-section resulting in different fitting values.  Without form factor, smaller $M_{S}$~(larger $\alpha'$) is required to compensate for the contribution from the $\Lambda$.

The fitting results demonstrate the universality of the string mass, $M_S$ for the OSS interaction in the elastic $pp, p\bar{p}$ scattering in the Regge limit~(the low energy scattering e.g. at $\sqrt{s}=546$ GeV gives a slightly different value). While the slope of the log-linear plot is determined by $M_{S}$ and $s$, the vertical intercept is determined also by strength of the coupling $g_{S}^{2}T^{2}$ in addition to the $M_{S}$~(from Eq.~(\ref{slope-open}) and (\ref{intercept-open})). The fitting values of the string coupling are alarmingly large and non-perturbative.  Using holographic model we could possibly attribute a fraction of the large value of $g_{S}$ to the warp factor and integral of the overlapping wave function in the holographic coordinate in order to keep the ``bare'' value in the perturbative regime.  However, in the next Section it turns out that once the closed string pomeron is included, the fitting values of the coupling reduce to fall in the perturbative regime and almost unchanged with respect to the collision energies, implying that both open and closed string pomerons must coexist in a consistent framework.

\subsection{Inclusion of the closed-string interaction }

A fascinating aspect of the string model is the open-closed string duality that connects loop diagrams of open string to the existence of closed string diagram by the worldsheet duality.  It is thus inevitable to include $t$-channel closed string diagram with vacuum quantum number together with the contributions from the open-string singlet.

The closed-string pomeron contribution to the $pp, p\bar{p}$ scattering has been explored in Ref. \cite{Domokos:2009hm}. Massive graviton~(closed-string) is coupled to the stress tensor of the protons (fermions) and identified as the spin-2 glueball state leading to the pomeron Regge trajectory. The holographic Sakai-Sugimoto~(SS) model~\cite{Sakai:2004cn} is used to calculate the relevant parameters and compared with fitting values from the experimental data. The differential cross-section of the closed-string pomeron contribution from Ref. \cite{Domokos:2009hm} is given by
\begin{eqnarray}
\frac{d\sigma}{dt} &=& \frac{\lambda^4}{16\pi}\left(\frac{\Gamma(-\chi)\Gamma(1-\alpha_c(t)/2)}{\Gamma(\alpha_c(t)/2-1-\chi)}\right)^2
\nonumber\\
&&\times \left( \frac{\alpha'_c s}{2}\right)^{2\alpha_c(t)-2} A^4(t)\,,
\label{harvey-diff-cross}
\end{eqnarray}
where $\lambda$ is the (graviton) glueball-proton-proton coupling constant which has dimension [mass]$^{-1}$ and
\begin{eqnarray}
\chi &=& 2 \alpha'_c\,m^2 + 3\alpha_c(0)/2 - 3 \,,\quad \alpha_c(t) = \alpha_c(0) + \alpha'_c\,t\,,
\nonumber\\
A(t) &=& \left(\frac{1}{1-t/\Lambda^2}\right)^2 \,.
\end{eqnarray}
Here $m$ is proton mass and $\Lambda$ is the dipole mass. In addition, the glueball mass, $m_g$ is defined from the Regge trajectory $(J = \alpha_c(0) + \alpha'_c t)$ by,
\begin{eqnarray}
m_g^2 = \frac{2 - \alpha_c(0)}{\alpha'_c}\,,
\end{eqnarray}
for spin $J=2$ and at the pole $t=m_g^2$\,. There are four free parameters in this model, $\lambda$\,, $\alpha_c(0)$\,, $\alpha'_c$ and $\Lambda$\,. By using the SS-model, one finds~\cite{Domokos:2010ma}
\begin{eqnarray}
\lambda = 9.02~{\rm GeV}^{-1}, ~m_g = 1.49~{\rm GeV}, ~\Lambda = 1.17~{\rm GeV}\,.
\end{eqnarray}
We note that $\Lambda = 1.17~{\rm GeV}$ is estimated from the four-dimensional Skyrme model \cite{Cebulla:2007ei}. By using formula in Eq. (\ref{harvey-diff-cross}) and fitting with the E-710 collaboration scattering data at $\sqrt{s} = 1800$ GeV, we obtain,
\begin{eqnarray}
\lambda &=& 8.62~{\rm GeV}^{-1},~\quad \alpha_c(0) = 1.074\,,
\nonumber\\
\alpha'_c &=& 0.286~{\rm GeV}^{-2},~\quad \Lambda = 0.970~{\rm GeV}\,.
\end{eqnarray}
The fitting values of $\lambda$ and other parameters are consistent with the results of the latest version of Ref. \cite{Domokos:2009hm} and Ref. \cite{Domokos:2010ma}. The value of $\alpha_{c}(0)\sim 1.06-1.08$ is also required in order to obtain a slowly increasing total cross section observed~\cite{Collins:1974en,Donnachie:1992ny}. Next, we consider the log-linear profile of the differential cross-section from the closed-string interaction. One finds,
\begin{eqnarray}
\log\left(\frac{d\sigma}{dt}\right) &=& \log\Bigg(\frac{\lambda^4}{16\pi}\left(\frac{\Gamma(-\chi)\Gamma(1-\alpha_c(t)/2)}{\Gamma(\alpha_c(t)/2-1-\chi)}\right)^2
\nonumber\\
&&\times \left( \frac{\alpha'_c s}{2}\right)^{2\alpha_c(t)-2} A^4(t) \Bigg)
\nonumber\\
&\approx& m_2t + C_2 \,,
\end{eqnarray}
where
\begin{eqnarray}
m_2 &=& 2\alpha_c'\log\left( \frac{\alpha'_c s}{2}\right) + \frac{8}{\Lambda^2} - \alpha_c'\psi^{(0)}\big( 1 - \alpha_c(0)/2 \big)
\nonumber\\
&&
-\, \alpha_c'\psi^{(0)}\big(\alpha_c(0)/2 - \chi - 1 \big),
\label{slope-close}
\\
C_2 &=& 2(\alpha_{c}(0)-1)\log\left(\frac{\alpha'_{c}s}{2}\right) \nonumber\\
&&+ 2\log\left(\frac{\lambda^2}{4\sqrt{\pi}}\frac{\Gamma(-\chi) \Gamma\left(1 -\alpha_c(0)/2 \right)}{\Gamma\left(\alpha_c(0)/2 - \chi -1 \right)}\right)
\label{intercept-close}
\end{eqnarray}
where $\psi^{(0)}(x)\equiv \displaystyle{\frac{1}{\Gamma(x)}\frac{d\Gamma(x)}{dx}}$ is the polygamma function.  Once again, the slope of log-linear of differential cross-section in the Regge limit for $\Lambda \gg 1$ GeV is given by
\begin{eqnarray}
m_2 &\approx& 2\alpha_c'\log\left( \frac{\alpha'_c s}{2}\right) .
\end{eqnarray}
Similar to the OSS amplitude case, only $\alpha_c'$  and $s$ determine the slope of the logarithmic plot for large $\Lambda$ and yet the effect of $\Lambda\simeq 1$ GeV cannot be neglected at the collision energies under consideration.

Although, as mentioned in Ref. \cite{Domokos:2009hm}, the closed-string (pomeron) interaction gives quite good agreement between the SS-model predictions and fitting results from experimental data but the closed-string contribution is not the only pomeron in the $pp, p\bar{p}$ elastic scattering.  As demonstrated in Ref.~\cite{Cullen:2000ef} in the stringy toy model of $U(2)$ QED and $U(4)$ QCD and more generically in Ref.~\cite{pbthesis,Burikham:2006an}, open-string singlet interaction in the $2\to 2$ process also contribute~(more dominantly since the OSS leading diagram is proportional to $g_{S}$ while the closed string diagram is proportional to $g_{S}^{2}$) as a gauge-singlet interaction, i.e., as a pomeron. Ref.~\cite{Cullen:2000ef} estimates the graviton-proton-proton coupling parameter $\lambda$ in the order
\be
\frac{\lambda^{2}}{m_{g}^{2}} \simeq \frac{\pi^{2}}{32}\frac{g_{S}^{2}}{M_{S}^{4}},
\ee
i.e., the $\lambda$ is proportional to the string coupling $g_{S}$.
To extend and improve the scattering model on the theoretical side, one has to take both the open-string and closed string contributions into account of the scattering process. The central result of this work is to consider the $pp, p\bar{p}$ elastic scattering and demonstrate the validity of the model. In the following, we
recall amplitudes from both of OSS from Eq. (\ref{SGS-amp}) and closed string interactions in the Regge limit, they read,
\begin{eqnarray}
\mathcal{A}_{\rm OSS} &=& \frac{1}{4}\,g_S\,T\,\frac{\alpha'}{t}\,\Gamma(1-\alpha' t)\left(1- e^{-i\pi\alpha't}\right)(\alpha' s)^{\alpha' t - 1}
\nonumber\\
&\times&\Big[ t\Big( 4\,\bar u(p_4)\,\bar u(p_1)\,\bar u(p_3)\,u(p_2)
\nonumber\\
&& \qquad\,-\, 2\,\bar u(p_4)\,\gamma_\mu\,\gamma_5\,u(p_1)\,\bar u(p_3)\,\gamma^\mu\,\gamma_5\,u(p_2)
\nonumber\\
&& \qquad\,-\,2\,\bar u(p_4)\,\gamma_\mu\,u(p_1)\,\bar u(p_3)\,\gamma^\mu\,u(p_2)
\nonumber\\
&& \qquad\, -\, 4\,\bar u(p_4) \,\gamma_5\,u(p_1)\,\bar u(p_3)\, \gamma_5\,u(p_2)  \Big)
\nonumber\\
&& \,-\, u\left( 3\leftrightarrow4 \right) \Big]\,A^2(t)
\\
\mathcal{A}_{\rm closed} &=& \frac{1}{8}\,\lambda^2\,\frac{\Gamma[-\chi]\,\Gamma[1-\alpha_c(t)/2]}{\Gamma[\alpha_c(t)/2-1-\chi]}\,
\frac{\alpha'_c}{2}
\nonumber\\
&& \times\,e^{-i\,\pi\,\alpha_c(t)/2} \left( \frac{\alpha'_c s}{2}\right)^{\alpha_c(t)-2}
\nonumber\\
& \times& \Big[ 2\,s\,\bar u(p_4)\,\gamma_\mu\,u(p_1)\,\bar u(p_3)\,\gamma^\mu\,u(p_2)
\label{inv-amp-close}\\
&&\; -\, 4\, p_1^\mu\,p_2^\nu\,\bar u(p_4)\,\gamma_\nu\,u(p_1)\,\bar u(p_3)\,\gamma_\mu\,u(p_2)\Big]\,A^2(t)\,,
\nonumber
\end{eqnarray}
where the $u(p)$ is standard positive energy Dirac spinor with normalization $\bar u(p)\,u(p) = 2\,m$ and the $\mathcal{A}_{\rm closed}$ is the closed string amplitude consistent with Ref. \cite{Domokos:2009hm}. The absolute square of the total amplitude, i.e. $\mathcal{A}_{\rm tot} = \mathcal{A}_{\rm OSS} + \mathcal{A}_{\rm closed}$ at the leading order terms of $s$ is given by
\begin{eqnarray}
\sum_{\rm spin}\Big|\mathcal{A}_{\rm tot} \Big|^2 &=& \sum_{\rm spin} \Big|\mathcal{A}_{\rm OSS} + \mathcal{A}_{\rm close}\Big|^2
\nonumber\\
&=&  32\, g_S^2\, T^2 \,(1-\cos(\pi\alpha't))
\nonumber\\
&\times& \Bigg( \frac{\Gamma(1-\alpha't)}{\alpha't}\Bigg)^2(\alpha's)^{2\alpha't + 2}\,A^4(t)
\nonumber\\
&+&\, 4\,s^4\,\lambda^4\left(\frac{\Gamma(-\chi)\,\Gamma(1-\alpha_c(t)/2)}{\Gamma(\alpha_c(t)/2-1-\chi)}\right)^2
\nonumber\\
&\;&\times \,\frac{\alpha_c'^2}{4}\left( \frac{\alpha'_c s}{2}\right)^{2\alpha_c(t)-4} A^4(t)
\nonumber\\
&+& 8\,g_S\,T\,\lambda^2\,s^2\left(\frac{\Gamma(-\chi)\,\Gamma(1-\alpha_c(t)/2)}{\Gamma(\alpha_c(t)/2-1-\chi)}\right)
\nonumber\\
&\;&\times \frac{\alpha_c'}{2}\left( \frac{\Gamma(1-\alpha't)}{\alpha't}\right) (\alpha' s)^{\alpha' t + 1}
\left( \frac{\alpha'_c s}{2}\right)^{\alpha_c(t)-2}
\nonumber\\
&\;&\times\Bigg( (1- e^{i\pi\alpha't})\,e^{-i\pi\alpha_c(t)/2}+ {\rm c.c.}\Bigg)A^4(t)
\label{total-amp}
\end{eqnarray}
And the differential cross-section for the total amplitude in Eq. (\ref{total-amp}) is given by,
\begin{eqnarray}
\frac{d\sigma}{dt} = \frac{1}{16\pi\,s^2}\left( \frac14 \sum_{\rm spin} \Big|\mathcal{A}_{\rm tot} \Big|^2\right)\,.
\label{diff-total}
\end{eqnarray}
This formula will be used for fitting the model parameters with the LHC-TOTEM \cite{Antchev:2018rec}, LHC-ATLAS \cite{Aad:2014dca}, D0 \cite{Abazov:2012qb}, E-710 \cite{Amos:1990fw} and CDF \cite{Abe:1993xx} scattering data in the small $(-t)$ region where single~(soft) pomeron exchange is dominant, see Figure~\ref{all-data}. Larger $(-t)$ scattering are complicated by other hard processes such as the hard~(BFKL) pomeron and perturbative QCD of the quarks and gluon. Notably, both soft and hard pomerons can be described by a single holographic string model in curved space~\cite{Brower:2006ea}.

\begin{figure*}[t]
\centering
\begin{tabular}{ccc}
\includegraphics[width=5cm,height=4cm,angle=0]{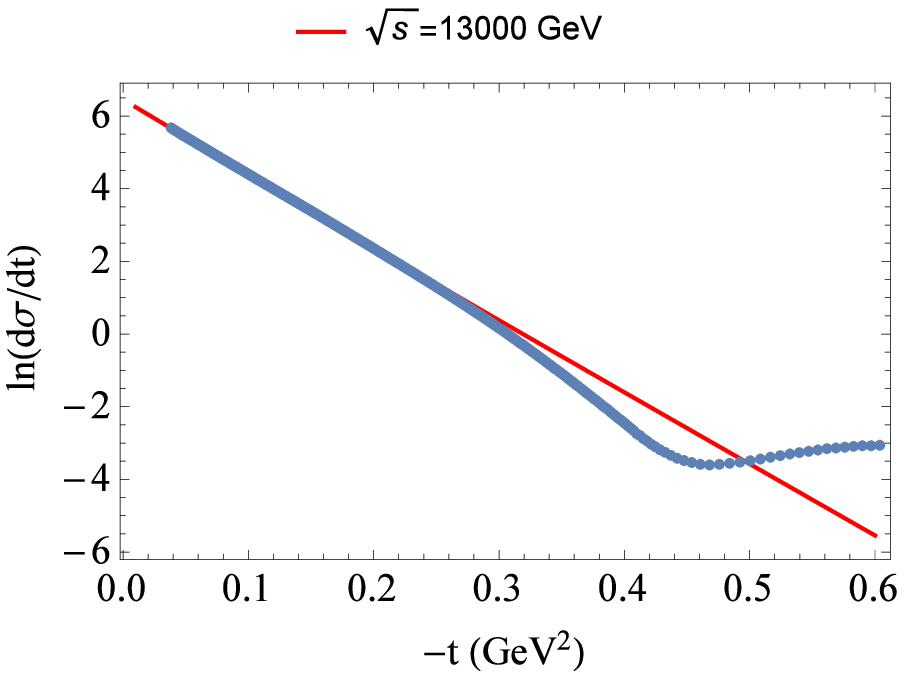}
&\includegraphics[width=5cm,height=4cm,angle=0]{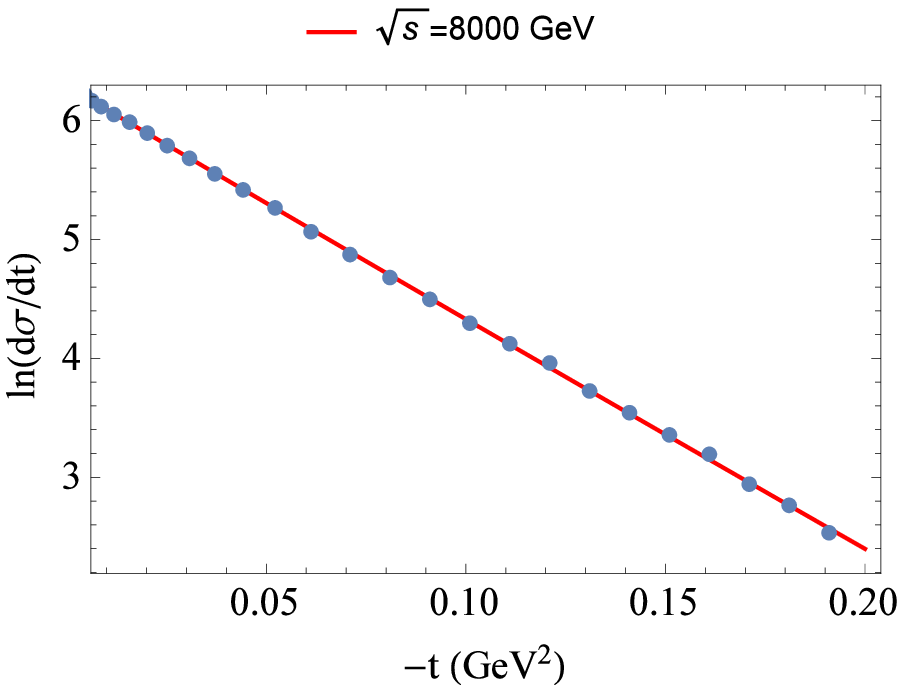}
&\includegraphics[width=5cm,height=4cm,angle=0]{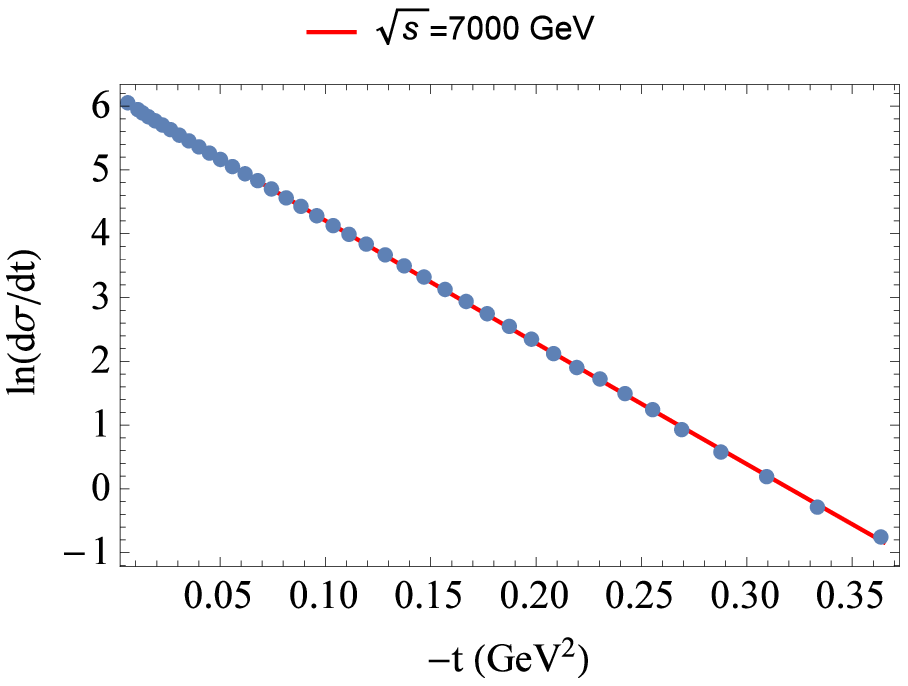}
\\
\includegraphics[width=5cm,height=4cm,angle=0]{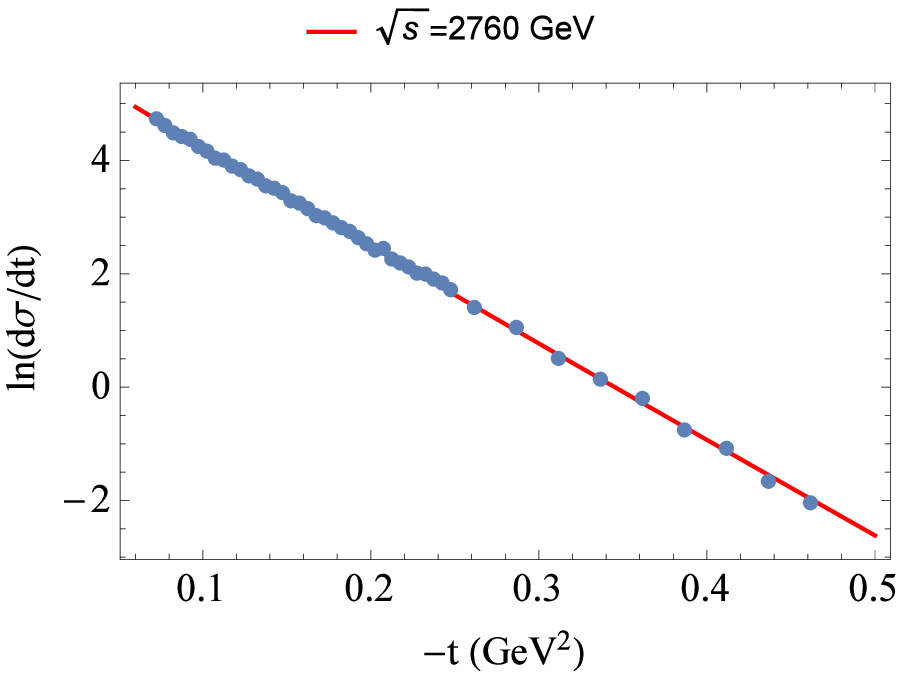}
&\includegraphics[width=5cm,height=4cm,angle=0]{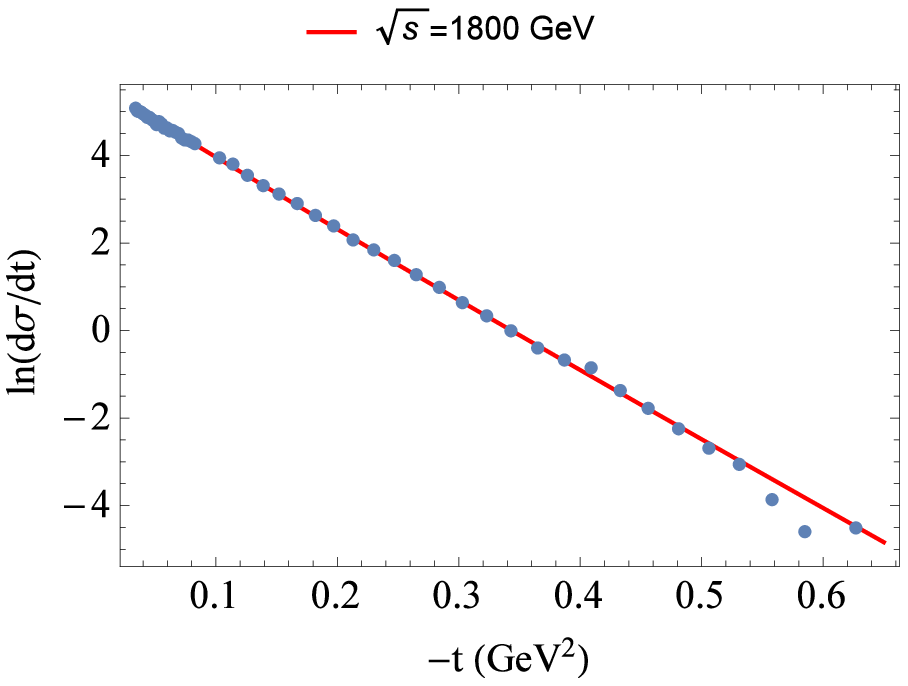}
&\includegraphics[width=5cm,height=4cm,angle=0]{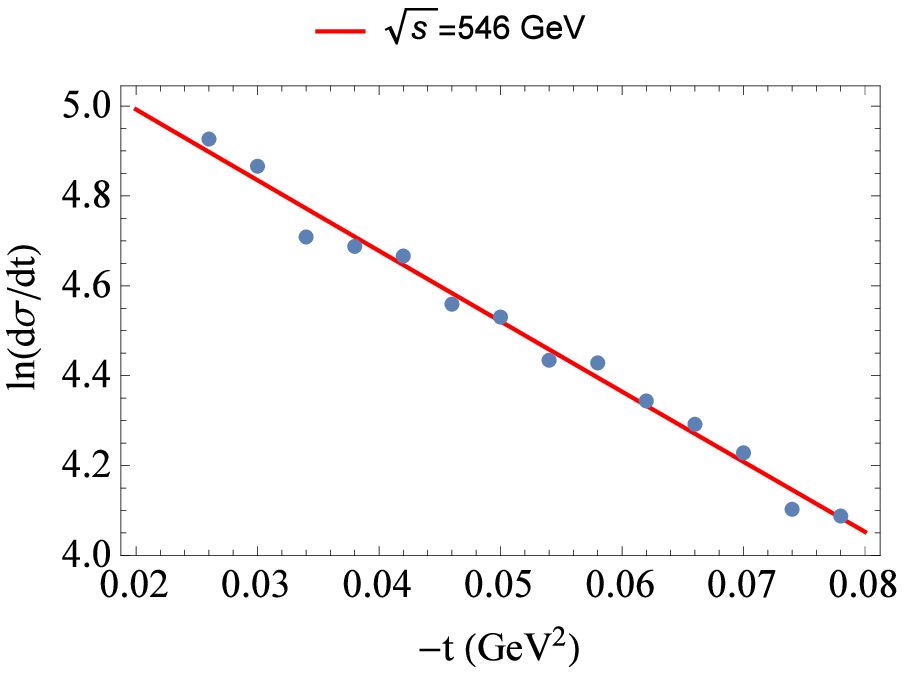}
\end{tabular}
\caption{The logarithmic fit of the $pp,p\bar{p}$ elastic collisions at $\sqrt{s}=13000, 8000,7000,2760,1800,546$ GeV by the combination of the open and closed string pomeron. The fits are performed with data in the small $(-t)$ region where linearity from single pomeron contribution is apparent.}\label{ffig}
\end{figure*}

According to the open-closed string duality, we set the string tension $\alpha' = \alpha_c'$. And fix $\Lambda=1.17$ GeV, $\alpha_{c}(0)=1.08$ in order to obtain slowly increasing total cross section $\sim s^{0.08}$~(we have also performed the fitting by taking $\alpha_{c}(0)$ as free parameter and the best-fit values are very close to $1.08$).  Since the closed string tree-level diagram is proportional to $g_{S}^{2}$, the graviton-proton-proton coupling is then rewritten using the string coupling as $\lambda=\alpha g_{S}$. The fitting results for various center of mass energies of the $pp,p\bar{p}$ collisions are given in Table~\ref{fittab} and Fig.~\ref{ffig}.

\begin{table}[h]
	\centering
	\begin{tabular}{|c|c|c|c|c|}
		\hline
		$\sqrt{s}/{\rm GeV}$ &~~$g_{S}$~~&$\alpha$/GeV$^{-1}$&~~ $M_{S}$/GeV ~~& $\lambda$/GeV$^{-1}$\\
		\hline
		13,000&1.39-10.94 &9.09-1.18& 1.65 &12.61-12.86\\
		8,000&1.37-10.86 &9.13-1.18&1.66 & 12.54-12.81 \\
		7,000&1.37-10.85&8.95-1.15& 1.66 &12.24-12.52 \\
		 2,760&1.48-11.77 &8.46-1.09&1.68&12.49-12.81 \\
		 1,960&1.64-13.18&7.42-0.95&1.68-1.69&12.15-12.48 \\
		 1,800&1.36-10.96&8.45-1.08&1.70-1.71&11.53-11.87 \\
		 546&1.34-10.61&9.04-1.18&1.66-1.67&12.08-12.48\\
		\hline
	\end{tabular}
	\caption{The best-fit string parameters from both open and closed string pomerons.}
	\label{fittab}
\end{table}
Remarkably, the fitting values of $M_{S}$ among all scattering data are very close to one another implying the universal value of the mass scale in the QCD-string model of the pomeron. Interestingly, the fitting values of the coupling $g_{S}$ could remain perturbative, i.e. $g_{S}/4\pi <1$, and yet take a wide range of values~(compensated by varying values of the corresponding $\alpha$) while the best-fit parameter of the effective closed string coupling hardly vary within the narrow range, $\lambda~(=\alpha g_{S})\simeq 12-13$ GeV$^{-1}$.

Moreover, the tensor glueball mass $m_g$ can be determined from the following formula \cite{Domokos:2009hm}
\begin{eqnarray}
m_g = M_{S}\sqrt{\left(2 - \alpha_c(0)\right)} \,.
\end{eqnarray}
All data universally gives $m_{g}\simeq1.6$ GeV. This glueball mass value is quite close to the estimation by the Sakai-Sugimoto model \cite{Domokos:2009hm}. However, this tensor glueball mass value is quite lower than lattice QCD estimation i.e. $m_g^{\rm lattice} = 2.40$ GeV~\cite{Morningstar:1999rf}. Interestingly, conformal symmetry breaking on one hand, generates mass of the glueball~(making graviton massive in the holographic picture) and at the same time shifts the intercept $\alpha_{c}(0)$ so that $J=2=\alpha_{c}(0)+\alpha' m_{g}^{2}$ is still preserved. The tiny mismatch between the string mass $M_{S}$ and the glueball mass $m_{g}~(<M_{S})$ results in a slightly larger than 1 value of the intercept $\alpha_{c}(0)=1.08$.

\subsection{Total cross-section from the OSS and closed string amplitudes}

The total cross-section of the $pp, p\bar{p}$ scattering from both the OSS and massive graviton amplitudes can be calculated by the optical theorem,
\begin{eqnarray}
\sigma_{\rm tot} = \frac{1}{s}\,{\rm Im}\Big(\mathcal{A}_{\rm OSS}(s, t)+\mathcal{A}_{\rm closed}(s, t)\Big)\Big|_{t=0}.
\label{optical}
\end{eqnarray}
By using the invariant amplitudes in Eqs. (\ref{sgsreg},\ref{closereg}), the total cross-section of the closed string and the OSS amplitudes reads,
\begin{eqnarray}
\sigma_{\rm tot} &=& \frac{\pi\,\lambda^2\,\Gamma[-\chi]\,\left( \alpha'_c s/2\right)^{\alpha_c(0)-1}}{\Gamma[\alpha_c(0)/2]\,\Gamma[\alpha_c(0)/2-1-\chi]}
\nonumber\\
 &&+\, \pi\, g_S\, T\,\alpha'\,,  \label{totsig}
\end{eqnarray}
where only $f_{\alpha}f_{\alpha}\to f_{\alpha}f_{\alpha}, \alpha=L, R$ contribute to the OSS amplitude at $t=0$ in the Regge limit. Notably, we obtain the constant total cross-section in the Regge limit i.e. $\sigma_{\rm tot}^{\rm OSS} \propto s^0$ for the open-string singlet contribution and almost constant $\sim s^{0.08}$ for $\alpha_{c}(0)=1.08$ from the closed-string amplitude. While the Regge theory gives general result of the total cross-section as $\sigma_{\rm tot} \propto s^{\alpha(0) - 1}$ where $\alpha(0)$ is the intercept of the angular momentum of any Regge trajectory. The pomeron trajectory with $\alpha(0) = 1$ is proposed by Chew~\cite{Chew:1961ev} and Gribov~\cite{Gribov:1961ex}. It has the vacuum quantum number, even signatures (charge and parity) and also satisfies Froissart bound i.e. $\sigma_{\rm tot} \leq (\ln s)^2$ at $s\to \infty$. The universality of the OSS is revealed by (\ref{totsig}), the effective size of each particle seen by another is proportional to the area of circle with radius equal to the string length $\ell_{S}\equiv \sqrt{\alpha'}$ and the string coupling $g_{S}$.

%%%%%%%%%%%%%%%%%%%%%%%%%%%%%%%%%%%%%%%%%%%%%%%%%%%%%%%%%%%%%%%%%%%%%%%%%%%%%%%%%%%%%%%%%%%%%%%%%%%%%%%%%%%%%%%%%%%%%%%%%%%%%%%%%%%%%%%%%%%%%%%%%%%%%%%%%%
\section{Discussions and conclusions}  \label{Disc}

The open and closed string pomerons are fit with the high energy $pp,p \bar{p}$ scattering in the Regge regime~(large $s$, small fixed $t$) at various energies. The fitting string parameters show remarkable universality in the values of the string mass scale $M_{S}$. The fitting coupling parameters $g_{S}$ and $\lambda$ also take almost universal values among scattering at various energies as shown in Table~\ref{fittab}. The fitting values of $g_{S}$ can vary in a wide range given that it is compensated by the closed-string coupling parameter $\alpha$ while keeping the product $\lambda=\alpha g_{S}$ mostly unchanged.

An insight obtained from the stringy picture of the pomeron is the relationship between the ratio of the two mass scales, $M_{S}=1/\sqrt{\alpha'}$ and $m_{g}$, and  the pomeron intercept $\alpha_{c}(0)$,
\begin{equation}
\alpha_{c}(0) = 2 - \frac{m_{g}^{2}}{M_{S}^{2}}.
\end{equation}
The slowly increasing total cross section $\sim\displaystyle{s^{1-(\frac{m_{g}}{M_{S}})^{2}}}$, requires that $m_{g}\lesssim M_{S}$. For $\alpha_{c}(0)=1.08$, the ratio is $m_{g}/M_{S}=0.96$. Since our fitting value $M_{S}\simeq 1.7$ GeV is less than the lattice value of $m_{g}^{\rm lattice}=2.4$ GeV, the string model is in tension with the lattice results. However, the lattice calculation ignored the light quarks and meson mixing so the actual pole mass value of the $2^{++}$ glueball could be quite different. The fitting value of $M_{S}$ also depends on the choice of the form factor scale $\Lambda$. For $\Lambda=0.94$ GeV~(proton mass), the fitting value becomes $M_{S}\simeq 1.8$ GeV~(still relatively universal among all data). Taken as a prediction, the string model prefers glueball pole mass around $1.6-1.7$ GeV. Remarkably, this lies within the values of various $2^{++}$~(``meson'') candidates observed in experiments, see Table \ref{gluecan}. With the same quantum numbers, glueball will mix with mesons as mass eigenstates listed among the $2^{++}$ candidates.

\begin{table}
\centering
\begin{tabular}{|c|c|c|}
\hline
$2^{++}$ meson & Mass (GeV) & width (GeV)\\
\hline
$f_2$(1270) & 1.276 & 0.190\\
$f_2$(1430) & 1.430 & undetermined\\
$f_2'$(1525) & 1.525 & 0.073\\
$f_2$(1565) & 1.562 & 0.134\\
$f_2$(1640) & 1.639 & 0.100\\
$f_2$(1810) & 1.815 & 0.200\\
$f_2$(1910) & 1.910 & 0.160\\
$f_2$(2010) & 2.011 & 0.200\\
$f_2$(2150) & 2.160 & 0.150\\
$f_2$(2220) & 2.231 & 0.023\\
$f_2$(2300) & 2.300 & 0.150\\
$f_2$(2340) & 2.345 & 0.322\\
\hline
\end{tabular}\caption{List of light $J^{PC}=2^{++}$ and $I=0$ mesons as glueball candidate \cite{Tanabashi:2018oca}.}
 \label{gluecan}
\end{table}

From the amplitudes in the Regge limit of the OSS and the closed string formulae (\ref{sgsreg}) and (\ref{closereg}), we can experimentally verify the existence of the OSS contribution by considering the scattering with polarized beams of proton and/or antiproton.  The closed string pomeron will give the same cross sections for any polarization while the OSS will give different results according to Eq.~(\ref{sgsreg}).

\section*{Acknowledgements}
P.B. is supported in part by the Thailand Research Fund (TRF),
Office of Higher Education Commission (OHEC) and Chulalongkorn University under grant RSA6180002.  D.S. is supported by Rachadapisek Sompote Fund for Postdoctoral Fellowship, Chulalongkorn University.
%%%%%%%%%%%%%%%%%%%%%%%%%%%%%%%%%%%%%%%%%%%%%%%%%%%%%%%%%%%%%%%%%%%%%%%%%%%%%%%%%%%%%%%%%%%%%%%%%%%%%%%%%%%%%%%%%%%%%%%%%%%%%%%%%%%%%%%%%%%%%%%%%%%%%%%%%%
\appendix
\section{helicity amplitudes of open and closed string pomerons in the Regge limit}  \label{app}

In this appendix, we present the helicity amplitude of the proton-proton scattering for OSS and close string amplitudes. The $A_{\rm OSS}$ amplitudes are given by,
\begin{eqnarray}
A_{\rm OSS} = \,g_S\,T\,K_{\rm OSS}(s,t,u)\,f(s,t,u) A^2(t),
\end{eqnarray}
where $K_{\rm OSS}(s,t,u)$ is kinematics function and characterizes the helicity structures of the amplitudes. It is given by
\begin{eqnarray}
K_{\rm OSS}(s,t,u)=
\begin{cases}
-\displaystyle{\frac{2\,s}{u\,t}}\,\quad{\rm for}\,~ f_\alpha\,f_\alpha \to f_\alpha\,f_\alpha\,,
\\
\\
-\displaystyle{\frac{2\,t}{s\,u}}\,\quad{\rm for}\,~ f_\alpha\,f_\beta \to f_\alpha\,f_\beta\,,
\end{cases}\label{sgsamp}
\end{eqnarray}
where $\alpha,\,\beta = L,R$, the helicity of the fermion. The amplitudes for $f\bar{f}\to f\bar{f}$ can be obtained by crossing $s\leftrightarrow t$.  Observe that in crossing, each helicity combination of (\ref{sgsamp}) turn into one another and thus the OSS differential cross section~(containing all helicity combinations) between $ff$ and $f\bar{f}$ are naturally {\it identical}.

In the Regge limit i.e., $u\,\to\,-s$\,, we find,
\begin{eqnarray}
A_{\rm OSS} =
\begin{cases}
2g_{S}T\displaystyle{\frac{1}{t}}f(s,t,u)A^2(t)~~{\rm for}\,~ f_\alpha\,f_\alpha \to f_\alpha f_\alpha,
\\
\\
2g_{S}T\displaystyle{\frac{t}{s^{2}}}f(s,t,u)A^2(t)~~{\rm for}\,~ f_\alpha\,f_\beta \to f_\alpha f_\beta,
\end{cases}\label{sgsreg}
\end{eqnarray}
where the function $f(s,t,u)$ in the Regge limit is given by Eq. (\ref{fstu-regge})\,. This means that different structures of the helicity amplitudes have different forms in the Regge limit. For the closed-string amplitudes in Eq. (\ref{inv-amp-close}), we obtain,
\begin{eqnarray}
A_{\rm closed} = \lambda^2\,K_{\rm closed}(s,t,u)\,\mathcal{P}_g\,A^2(t),
\end{eqnarray}
where $K_{\rm closed}(s,t,u)$ is the helicity dependent kinematic function and
\begin{equation}
\mathcal{P}_g =\frac{\Gamma[-\chi]\,\Gamma[1-\alpha_c(t)/2]}{\Gamma[\alpha_c(t)/2-1-\chi]}(\alpha'_c/2)e^{-i\,\pi\,\alpha_c(t)/2}\left( \alpha'_c s/2\right)^{\alpha_c(t)-2}.
\end{equation}
The $K_{\rm closed}(s,t,u)$ function in the Regge limit can be written as
\begin{eqnarray}
K_{\rm closed}(s,t,u) =
\begin{cases}
\frac12\,s^2\left(1-\displaystyle{\frac{u}{s}}\right)\quad{\rm for}\,~ f_\alpha\,f_\alpha \to f_\alpha\,f_\alpha\,,
\\
\\
-s\,u\,\quad{\rm for}\,~ f_\alpha\,f_\beta \to f_\alpha\,f_\beta\,.
\end{cases}
\end{eqnarray}
In the Regge limit i.e., $u\approx -s$\,, one finds,
\begin{eqnarray}
A_{\rm closed} =
\begin{cases}
\lambda^2\,s^2\,\mathcal{P}_g A^2(t)\quad{\rm for}\,~ f_\alpha\,f_\alpha \to f_\alpha\,f_\alpha\,,
\\
\\
\lambda^2\,s^2\,\mathcal{P}_g A^2(t)\quad{\rm for}\,~ f_\alpha\,f_\beta \to f_\alpha\,f_\beta\,.\\
\end{cases}\label{closereg}
\end{eqnarray}
The helicity amplitudes of the $A_{\rm closed}$ have the same form for all helicity configurations in the Regge limit.

%%%%%%%%%%%%%%%%%%%%%%%%%%%%%%%%%%%%%%%%%%%%%%%%%%%%%%%%%%%%%%%%%%%%%%%%%%%%%%%%%%%%%%%%%%%%%%%%%%%%%%%%%%%%%%%%%%%%%%%%%%%%%%%%%%%%%%%%%%%%%%%%%%%%%%%%%%


\begin{thebibliography}{99}

\bibitem{Nambu:1969se}
  Y.~Nambu,
  ``Quark model and the factorization of the Veneziano amplitude,''
  In *Detroit 1969, Proceedings, International Conference on Symmetries and Quark Models, Gordon and Breach, pp.269-277.
  %2 citations counted in INSPIRE as of 20 Nov 2018

\bibitem{Koba:1969rw}
  Z.~Koba and H.~B.~Nielsen,
  %``Reaction amplitude for n mesons: A Generalization of the Veneziano-Bardakci-Ruegg-Virasora model,''
  Nucl.\ Phys.\ B {\bf 10}, 633 (1969).
 % doi:10.1016/0550-3213(69)90331-9
  %%CITATION = doi:10.1016/0550-3213(69)90331-9;%%
  %209 citations counted in INSPIRE as of 20 Nov 2018

\bibitem{Susskind:1970qz}
  L.~Susskind,
  %``Structure of hadrons implied by duality,''
  Phys.\ Rev.\ D {\bf 1}, 1182 (1970).
  %doi:10.1103/PhysRevD.1.1182
  %%CITATION = doi:10.1103/PhysRevD.1.1182;%%
  %119 citations counted in INSPIRE as of 20 Nov 2018


\bibitem{Tanabashi:2018oca}
  M.~Tanabashi {\it et al.} [Particle Data Group],
  %``Review of Particle Physics,''
  Phys.\ Rev.\ D {\bf 98}, no. 3, 030001 (2018).
  %doi:10.1103/PhysRevD.98.030001
  %%CITATION = doi:10.1103/PhysRevD.98.030001;%%
  %480 citations counted in INSPIRE as of 20 Nov 2018


\bibitem{Yoneya:1974jg}
  T.~Yoneya,
  %``Connection of Dual Models to Electrodynamics and Gravidynamics,''
  Prog.\ Theor.\ Phys.\  {\bf 51}, 1907 (1974).
  %doi:10.1143/PTP.51.1907

\bibitem{Scherk:1974ca}
  J.~Scherk and J.~H.~Schwarz,
  %``Dual Models for Nonhadrons,''
  Nucl.\ Phys.\ B {\bf 81}, 118 (1974).
  %doi:10.1016/0550-3213(74)90010-8


\bibitem{Green:1987sp}
  M.~B.~Green, J.~H.~Schwarz and E.~Witten,
  Superstring Theory, Vol. 1 and 2, Cambridge university press (1987).
  %281 citations counted in INSPIRE as of 09 Oct 2018


\bibitem{Donnachie:1992ny}
  A.~Donnachie and P.~V.~Landshoff,
  %``Total cross-sections,''
  Phys.\ Lett.\ B {\bf 296}, 227 (1992)
  %doi:10.1016/0370-2693(92)90832-O
  [hep-ph/9209205].
  %%CITATION = doi:10.1016/0370-2693(92)90832-O;%%
  %1261 citations counted in INSPIRE as of 20 Nov 2018

\bibitem{Ewerz:2016onn}
  C.~Ewerz, P.~Lebiedowicz, O.~Nachtmann and A.~Szczurek,
  %``Helicity in proton–proton elastic scattering and the spin structure of the pomeron,''
  Phys.\ Lett.\ B {\bf 763}, 382 (2016)
  %doi:10.1016/j.physletb.2016.10.064
  [arXiv:1606.08067 [hep-ph]].
  %%CITATION = doi:10.1016/j.physletb.2016.10.064;%%
  %12 citations counted in INSPIRE as of 20 Nov 2018

\bibitem{Ochs:2013gi}
  W.~Ochs,
  %``The Status of Glueballs,''
  J.\ Phys.\ G {\bf 40}, 043001 (2013)
  %doi:10.1088/0954-3899/40/4/043001
  [arXiv:1301.5183 [hep-ph]].
  %%CITATION = doi:10.1088/0954-3899/40/4/043001;%%
  %137 citations counted in INSPIRE as of 20 Nov 2018


\bibitem{Maldacena:1997re}
  J.~M.~Maldacena,
  %``The Large N limit of superconformal field theories and supergravity,''
  Int.\ J.\ Theor.\ Phys.\  {\bf 38}, 1113 (1999)
  [Adv.\ Theor.\ Math.\ Phys.\  {\bf 2}, 231 (1998)]
  %doi:10.1023/A:1026654312961, 10.4310/ATMP.1998.v2.n2.a1
  [hep-th/9711200].
  %%CITATION = doi:10.1023/A:1026654312961, 10.4310/ATMP.1998.v2.n2.a1;%%
  %14058 citations counted in INSPIRE as of 13 Oct 2018


\bibitem{Witten:1998xy}
  E.~Witten,
  %``Baryons and branes in anti-de Sitter space,''
  JHEP {\bf 9807}, 006 (1998)
  %doi:10.1088/1126-6708/1998/07/006
  [hep-th/9805112].
  %%CITATION = doi:10.1088/1126-6708/1998/07/006;%%
  %575 citations counted in INSPIRE as of 20 Nov 2018

\bibitem{Gross:1998gk}
  D.~J.~Gross and H.~Ooguri,
  %``Aspects of large N gauge theory dynamics as seen by string theory,''
  Phys.\ Rev.\ D {\bf 58}, 106002 (1998)
  %doi:10.1103/PhysRevD.58.106002
  [hep-th/9805129].
  %%CITATION = doi:10.1103/PhysRevD.58.106002;%%
  %365 citations counted in INSPIRE as of 20 Nov 2018

\bibitem{Burikham:2008cr}
  P.~Burikham, A.~Chatrabhuti and E.~Hirunsirisawat,
  %``Exotic Multi-quark States in the Deconfined Phase from Gravity Dual Models,''
  JHEP {\bf 0905}, 006 (2009)
  doi:10.1088/1126-6708/2009/05/006
  [arXiv:0811.0243 [hep-ph]].

%\cite{Burikham:2006an}
\bibitem{Burikham:2006an}
  P.~Burikham,
  %``Universal open-string singlet interaction,''
  Int.\ J.\ Mod.\ Phys.\ A {\bf 22}, 29 (2007).
  %doi:10.1142/S0217751X07035264
  %%CITATION = doi:10.1142/S0217751X07035264;%%
  %5 citations counted in INSPIRE as of 22 Aug 2018

\bibitem{Cullen:2000ef}
  S.~Cullen, M.~Perelstein and M.~E.~Peskin,
  %``TeV strings and collider probes of large extra dimensions,''
  Phys.\ Rev.\ D {\bf 62}, 055012 (2000)
  %doi:10.1103/PhysRevD.62.055012
  [hep-ph/0001166].
  %%CITATION = doi:10.1103/PhysRevD.62.055012;%%
  %304 citations counted in INSPIRE as of 13 Oct 2018

\bibitem{pbthesis}
  P.~Burikham, PhD thesis,
  %``Physics beyond the standard model and collider phenomenology,''
  UMI-31-86286.
  %%CITATION = UMI-31-86286;%%


\bibitem{Burikham:2003ha}
  P.~Burikham, T.~Han, F.~Hussain and D.~W.~McKay,
  %``Bounds on four fermion contact interactions induced by string resonances,''
  Phys.\ Rev.\ D {\bf 69}, 095001 (2004)
  %doi:10.1103/PhysRevD.69.095001
  [hep-ph/0309132].
  %%CITATION = doi:10.1103/PhysRevD.69.095001;%%
  %20 citations counted in INSPIRE as of 13 Oct 2018

\bibitem{Burikham:2004su}
  P.~Burikham, T.~Figy and T.~Han,
  %``TeV-scale string resonances at hadron colliders,''
  Phys.\ Rev.\ D {\bf 71}, 016005 (2005)
  Erratum: [Phys.\ Rev.\ D {\bf 71}, 019905 (2005)]
 % doi:10.1103/PhysRevD.71.016005, 10.1103/PhysRevD.71.019905
  [hep-ph/0411094].
  %%CITATION = doi:10.1103/PhysRevD.71.016005, 10.1103/PhysRevD.71.019905;%%
  %53 citations counted in INSPIRE as of 13 Oct 2018

\bibitem{Burikham:2004uu}
  P.~Burikham,
  %``Similarity between Kaluza-Klein and open-string amplitudes in diphoton production,''
  JHEP {\bf 0407}, 053 (2004)
  %doi:10.1088/1126-6708/2004/07/053
  [hep-ph/0407271].
  %%CITATION = doi:10.1088/1126-6708/2004/07/053;%%
  %5 citations counted in INSPIRE as of 13 Oct 2018

\bibitem{Burikham:2006hi}
  P.~Burikham,
  %``TeV-scale stringy signatures at the electron-positron collider,''
  Phys.\ Rev.\ D {\bf 73}, 055006 (2006)
  %doi:10.1103/PhysRevD.73.055006
  [hep-ph/0601142].
  %%CITATION = doi:10.1103/PhysRevD.73.055006;%%
  %2 citations

%\cite{Domokos:2009hm}
\bibitem{Domokos:2009hm}
  S.~K.~Domokos, J.~A.~Harvey and N.~Mann,
  %``The Pomeron contribution to p p and p anti-p scattering in AdS/QCD,''
  Phys.\ Rev.\ D {\bf 80}, 126015 (2009)
  %doi:10.1103/PhysRevD.80.126015
  [arXiv:0907.1084 [hep-ph]].
  %%CITATION = doi:10.1103/PhysRevD.80.126015;%%
  %24 citations counted in INSPIRE as of 21 Sep 2018

\bibitem{Domokos:2010ma}
  S.~K.~Domokos, J.~A.~Harvey and N.~Mann,
  %``Setting the scale of the p p and p bar p total cross sections using AdS/QCD,''
  Phys.\ Rev.\ D {\bf 82}, 106007 (2010)
  %doi:10.1103/PhysRevD.82.106007
  [arXiv:1008.2963 [hep-th]].
  %%CITATION = doi:10.1103/PhysRevD.82.106007;%%
  %9 citations counted in INSPIRE as of 16 Oct 2018

\bibitem{Avsar:2009hc}
  E.~Avsar, Y.~Hatta and T.~Matsuo,
  %``Odderon in baryon-baryon scattering from the AdS/CFT correspondence,''
  JHEP {\bf 1003}, 037 (2010)
  %doi:10.1007/JHEP03(2010)037
  [arXiv:0912.3806 [hep-th]].
  %%CITATION = doi:10.1007/JHEP03(2010)037;%%
  %9 citations counted in INSPIRE as of 16 Oct 2018


\bibitem{Anderson:2014jia}
  N.~Anderson, S.~K.~Domokos, J.~A.~Harvey and N.~Mann,
  %``Central production of $\eta$ and $\eta′$ via double Pomeron exchange in the Sakai-Sugimoto model,''
  Phys.\ Rev.\ D {\bf 90}, no. 8, 086010 (2014)
  %doi:10.1103/PhysRevD.90.086010
  [arXiv:1406.7010 [hep-ph]].
  %%CITATION = doi:10.1103/PhysRevD.90.086010;%%
  %9 citations counted in INSPIRE as of 16 Oct 2018

\bibitem{Iatrakis:2016rvj}
  I.~Iatrakis, A.~Ramamurti and E.~Shuryak,
  %``Pomeron Interactions from the Einstein-Hilbert Action,''
  Phys.\ Rev.\ D {\bf 94}, no. 4, 045005 (2016)
  %doi:10.1103/PhysRevD.94.045005
  [arXiv:1602.05014 [hep-ph]].
  %%CITATION = doi:10.1103/PhysRevD.94.045005;%%
  %6 citations counted in INSPIRE as of 16 Oct 2018

\bibitem{Anderson:2016zon}
  N.~Anderson, S.~Domokos and N.~Mann,
  %``Central production of $\eta$ via double Pomeron exchange and double Reggeon exchange in the Sakai-Sugimoto model,''
  Phys.\ Rev.\ D {\bf 96}, no. 4, 046002 (2017)
  %doi:10.1103/PhysRevD.96.046002
  [arXiv:1612.07457 [hep-ph]].
  %%CITATION = doi:10.1103/PhysRevD.96.046002;%%
  %3 citations counted in INSPIRE as of 16 Oct 2018

\bibitem{Hu:2017iix}
  Z.~Hu, B.~Maddock and N.~Mann,
  %``A Second Look at String-Inspired Models for Proton-Proton Scattering via Pomeron Exchange,''
  JHEP {\bf 1808}, 093 (2018)
  %doi:10.1007/JHEP08(2018)093
  [arXiv:1710.02463 [hep-ph]].
  %%CITATION = doi:10.1007/JHEP08(2018)093;%%

\bibitem{Xie:2019soz}
  W.~Xie, A.~Watanabe and M.~Huang,
  %``Elastic proton-proton scattering at LHC energies in holographic QCD,''
  arXiv:1901.09564 [hep-ph].
  %%CITATION = ARXIV:1901.09564;%%

\bibitem{Veneziano:1968yb}
  G.~Veneziano,
  %``Construction of a crossing - symmetric, Regge behaved amplitude for linearly rising trajectories,''
  Nuovo Cim.\ A {\bf 57}, 190 (1968).
  %doi:10.1007/BF02824451
  %%CITATION = doi:10.1007/BF02824451;%%
  %1342 citations counted in INSPIRE as of 13 Oct 2018

\bibitem{Itzykson:1980rh}
  C.~Itzykson and J.~B.~Zuber,
  %``Quantum Field Theory,''
  New York, Usa: Mcgraw-hill (1980), page 160-162, (International Series In Pure and Applied Physics).
  %342 citations counted in INSPIRE as of 10 Sep 2018

\bibitem{Frampton:1986wv}
  P.~H.~Frampton,
  ``Dual Resonance Models And Superstrings,''
  Singapore, Singapore: World Scientific (1986).
  %1 citations counted in INSPIRE as of 13 Oct 2018

\bibitem{Donnachie:2002en}
  S.~Donnachie, H.~G.~Dosch, O.~Nachtmann and P.~Landshoff,
  ``Pomeron physics and QCD,''
  Cambridge university press  (2002).
  %%CITATION = CMPCE,19,1;%%
  %119 citations counted in INSPIRE as of 16 Oct 2018

\bibitem{Wong:1995jf}
  C.~Y.~Wong,
  ``Introduction to high-energy heavy ion collisions,''
  Singapore, Singapore: World Scientific (1994).
  %16 citations counted in INSPIRE as of 13 Oct 2018

\bibitem{Pagels:1966zza}
  H.~Pagels,
  %``Energy-Momentum Structure Form Factors of Particles,''
  Phys.\ Rev.\  {\bf 144}, 1250 (1966).
  %doi:10.1103/PhysRev.144.1250

\bibitem{Shtabovenko:2016sxi}
  V.~Shtabovenko, R.~Mertig and F.~Orellana,
  %``New Developments in FeynCalc 9.0,''
  Comput.\ Phys.\ Commun.\  {\bf 207}, 432 (2016)
  %doi:10.1016/j.cpc.2016.06.008
  [arXiv:1601.01167 [hep-ph]].
  %%CITATION = doi:10.1016/j.cpc.2016.06.008;%%
  %145 citations counted in INSPIRE as of 22 Aug 2018

\bibitem{Mertig:1990an}
  R.~Mertig, M.~Bohm and A.~Denner,
  %``FEYN CALC: Computer algebraic calculation of Feynman amplitudes,''
  Comput.\ Phys.\ Commun.\  {\bf 64}, 345 (1991).
  %doi:10.1016/0010-4655(91)90130-D
  %%CITATION = doi:10.1016/0010-4655(91)90130-D;%%
  %709 citations counted in INSPIRE as of 22 Aug 2018

\bibitem{durham}
The Durham HEP database, http://durpdg.dur.ac.uk/ .

\bibitem{Amos:1990fw}
  N.~A.~Amos {\it et al.} [E-710 Collaboration],
  %``$\bar{p}p$ elastic scattering at $\sqrt{s}$ = 1.8-TeV from |t| = $0.034-GeV/c^{2}$ to $0.65-GeV/c^{2}$,''
  Phys.\ Lett.\ B {\bf 247}, 127 (1990).
  %doi:10.1016/0370-2693(90)91060-O
  %%CITATION = doi:10.1016/0370-2693(90)91060-O;%%
  %171 citations counted in INSPIRE as of 10 Sep 2018

\bibitem{Aad:2014dca}
  G.~Aad {\it et al.} [ATLAS Collaboration],
  %``Measurement of the total cross section from elastic scattering in pp collisions at $\sqrt{s}=7$ TeV with the ATLAS detector,''
  Nucl.\ Phys.\ B {\bf 889}, 486 (2014)
  %doi:10.1016/j.nuclphysb.2014.10.019
  [arXiv:1408.5778 [hep-ex]].
  %%CITATION = doi:10.1016/j.nuclphysb.2014.10.019;%%
  %137 citations counted in INSPIRE as of 20 Sep 2018

%\cite{Abazov:2012qb}
\bibitem{Abazov:2012qb}
  V.~M.~Abazov {\it et al.} [D0 Collaboration],
  %``Measurement of the differential cross section $d\sigma/dt$ in elastic $p\bar{p}$ scattering at $\sqrt{s}=1.96$ TeV,''
  Phys.\ Rev.\ D {\bf 86}, 012009 (2012)
  %doi:10.1103/PhysRevD.86.012009
  [arXiv:1206.0687 [hep-ex]].
  %%CITATION = doi:10.1103/PhysRevD.86.012009;%%
  %32 citations counted in INSPIRE as of 20 Sep 2018

\bibitem{Abe:1993xx}
  F.~Abe {\it et al.} [CDF Collaboration],
  %``Measurement of small angle $\bar{p}p$ elastic scattering at $\sqrt{s} = 546$ GeV and 1800 GeV,''
  Phys.\ Rev.\ D {\bf 50}, 5518 (1994).
 % doi:10.1103/PhysRevD.50.5518
  %%CITATION = doi:10.1103/PhysRevD.50.5518;%%
  %249 citations counted in INSPIRE as of 20 Sep 2018

\bibitem{Cebulla:2007ei}
  C.~Cebulla, K.~Goeke, J.~Ossmann and P.~Schweitzer,
  %``The Nucleon form-factors of the energy momentum tensor in the Skyrme model,''
  Nucl.\ Phys.\ A {\bf 794}, 87 (2007)
  %doi:10.1016/j.nuclphysa.2007.08.004
  [hep-ph/0703025 [hep-ph]].
  %%CITATION = doi:10.1016/j.nuclphysa.2007.08.004;%%
  %30 citations counted in INSPIRE as of 18 Sep 2018

\bibitem{Sakai:2004cn}
  T.~Sakai and S.~Sugimoto,
  %``Low energy hadron physics in holographic QCD,''
  Prog.\ Theor.\ Phys.\  {\bf 113}, 843 (2005)
  %doi:10.1143/PTP.113.843
  [hep-th/0412141].
  %%CITATION = doi:10.1143/PTP.113.843;%%
  %1172 citations counted in INSPIRE as of 13 Oct 2018

\bibitem{Collins:1974en}
  P.~D.~B.~Collins, F.~D.~Gault and A.~D.~Martin,
  %``Proton Proton Scattering and the Pomeron,''
  Nucl.\ Phys.\ B {\bf 80}, 135 (1974).
  %doi:10.1016/0550-3213(74)90035-2

%\cite{Antchev:2018rec}
\bibitem{Antchev:2018rec}
  G.~Antchev {\it et al.} [TOTEM Collaboration],
  %``Elastic differential cross-section ${\rm d}\sigma/{\rm d}t$ at $\sqrt{s}=$2.76 TeV and implications on the existence of a colourless 3-gluon bound state,''
  arXiv:1812.08610 [hep-ex];
  %%CITATION = ARXIV:1812.08610;%%
  %1 citations counted in INSPIRE as of 31 Jan 2019
  G.~Antchev {\it et al.} [TOTEM Collaboration],
  %``Measurement of elastic pp scattering at $\sqrt{\hbox {s}} = \hbox {8}$  TeV in the Coulomb–nuclear interference region: determination of the $\mathbf {\rho }$ -parameter and the total cross-section,''
  Eur.\ Phys.\ J.\ C {\bf 76}, no. 12, 661 (2016);
  %doi:10.1140/epjc/s10052-016-4399-8
  [arXiv:1610.00603 [nucl-ex]];
  %%CITATION = doi:10.1140/epjc/s10052-016-4399-8;%%
  %73 citations counted in INSPIRE as of 31 Jan 2019
  G.~Antchev {\it et al.} [TOTEM Collaboration],
  %``Elastic differential cross-section measurement at $\sqrt{s}=13$ TeV by TOTEM,''
  arXiv:1812.08283 [hep-ex].

\bibitem{Brower:2006ea}
  R.~C.~Brower, J.~Polchinski, M.~J.~Strassler and C.~I.~Tan,
  %``The Pomeron and gauge/string duality,''
  JHEP {\bf 0712}, 005 (2007)
  %doi:10.1088/1126-6708/2007/12/005
  [hep-th/0603115].

\bibitem{Morningstar:1999rf}
  C.~J.~Morningstar and M.~J.~Peardon,
  %``The Glueball spectrum from an anisotropic lattice study,''
  Phys.\ Rev.\ D {\bf 60}, 034509 (1999)
  %doi:10.1103/PhysRevD.60.034509
  [hep-lat/9901004].
  %%CITATION = doi:10.1103/PhysRevD.60.034509;%%
  %832 citations counted in INSPIRE as of 01 Oct 2018

\bibitem{Chew:1961ev}
  G.~F.~Chew and S.~C.~Frautschi,
  %``Principle of Equivalence for All Strongly Interacting Particles Within the S Matrix Framework,''
  Phys.\ Rev.\ Lett.\  {\bf 7}, 394 (1961).
  %doi:10.1103/PhysRevLett.7.394
  %%CITATION = doi:10.1103/PhysRevLett.7.394;%%
  %315 citations counted in INSPIRE as of 20 Nov 2018

\bibitem{Gribov:1961ex}
  V.~N.~Gribov,
  %``Possible Asymptotic Behavior of Elastic Scattering,''
  JETP \  {\bf 14} (1962) 478.
  %%CITATION = JTPLA,41,667;%%
  %35 citations counted in INSPIRE as of 20 Nov 2018
%%\cite{Brower:2006ea}
%\bibitem{Brower:2006ea}
%  R.~C.~Brower, J.~Polchinski, M.~J.~Strassler and C.~I.~Tan,
%  ``The Pomeron and gauge/string duality,''
%  JHEP {\bf 0712}, 005 (2007)
%  %doi:10.1088/1126-6708/2007/12/005
%  [hep-th/0603115].
%  %%CITATION = doi:10.1088/1126-6708/2007/12/005;%%
%  %307 citations counted in INSPIRE as of 01 Jun 2018
 \end{thebibliography}
\end{document}